\pdfoutput=1
\documentclass{JINST}
\usepackage{graphicx}
\usepackage[figuresright]{rotating}

\title{Mechanical Design and Material Budget of the CMS Barrel Pixel Detector}

\author{C.~Amsler$^a$, K.~B\"osiger$^a$, V.~Chiochia$^a$, W.~Erdmann$^b$, K.~Gabathuler$^b$, R.~Horisberger$^b$, S.~K\"onig$^b$, D.~Kotlinski$^b$, R.~Maier$^a$, B.~Meier$^b$, Hp.~Meyer$^a$, A.~Rizzi$^c$, P.~Robmann$^a$, S.~Scherr$^a$, A.~Schmidt$^a$\thanks{Corresponding author.}, S.~Steiner$^a$, S.~Streuli$^b$\\
\llap{$^a$}Universit\"at Z\"urich, Physik-Institut,\\
  Winterthurerstr. 190, CH-8057 Z\"urich, Switzerland\\
\llap{$^b$}Paul Scherrer Institut,\\
  CH-5232 Villigen, Switzerland\\
  \llap{$^c$}ETH Z\"urich, Institute for Particle Physics,\\
  CH-8093 Z\"urich, Switzerland\\
  E-mail: \email{Alexander.Schmidt@cern.ch}}

\abstract{The Compact Muon Solenoid experiment at the Large Hadron Collider at CERN includes a silicon pixel detector as its innermost component. Its main task is the precise reconstruction of charged particles  close to the primary interaction vertex. This paper gives an overview of the mechanical requirements and design choices for the barrel pixel detector. The distribution of material in the detector as well as its description in the Monte Carlo simulation are discussed in detail.}


\begin{document}

 \section{Introduction}
\label{sec:introduction}
At the LHC design luminosity of $10^{34} \ \rm cm^{-2}s^{-1}$, about 1000 particles will be traversing the tracker for each bunch crossing every $\rm 25 \ ns$. At a radius of $\rm 4 \ cm$ this corresponds to a hit rate density of $\rm 1 \ MHz/mm^2$.  The tracker is required to operate in this high radiation environment with a reasonable lifetime of several years. The trajectories of charged particles need to be measured efficiently and with high precision. The amount of material needs to be kept to a minimum in order to limit effects such as multiple scattering, bremsstrahlung, photon conversions and nuclear interactions. These requirements and limitations strongly suggested an all-silicon design. 

The CMS tracking detector consists of $\rm 200 \ m^2$ of active silicon. It is therefore the largest silicon tracking device ever built \cite{bib:detectorPaper,bib:trackerTDR,bib:trackerTDRAddendum}. 

%The outer part consists of 15148 silicon micro-strip detector modules arranged in 10 barrel layers, extending to a radius of 1.1~m, and 9 endcap disks on each side which complete the acceptance up to a pseudo-rapidity of $|\eta| < 2.5$.

The pixel detector as the innermost part of CMS consists of three barrel layers at radii of $\rm 4.4 $, $7.4$ and $\rm 10.2 \ cm$, respectively, with a length of $\rm 53 \ cm$. It is completed by two forward disks on each side, located along the beam axis at $z=\pm 34.5 \ {\rm cm}$ and $z=\pm 46.5 \ {\rm cm}$, extending in radius from $6 \ \rm cm$ to $15 \ \rm cm$. The pixel detector has 66 million pixels in total, of which 48 millions belong to the barrel.

Details on the mechanical design of the pixel barrel detector (PXB)  are given in Section~\ref{sec:Design}, while the material budget is discussed in Section~\ref{sec:MaterialBudget}.
The description of the detector in the Monte Carlo simulation is outlined in Section~\ref{sec:MCsimulation}.

\section{Detector Design\label{sec:Design}}
For the purpose of the following discussion the detector is subdivided into three logical parts: the central barrel (containing the detector modules), the endflange and the supply tube. 

\begin{figure}[htbp!]
\begin{center}
\includegraphics[width=0.75\textwidth]{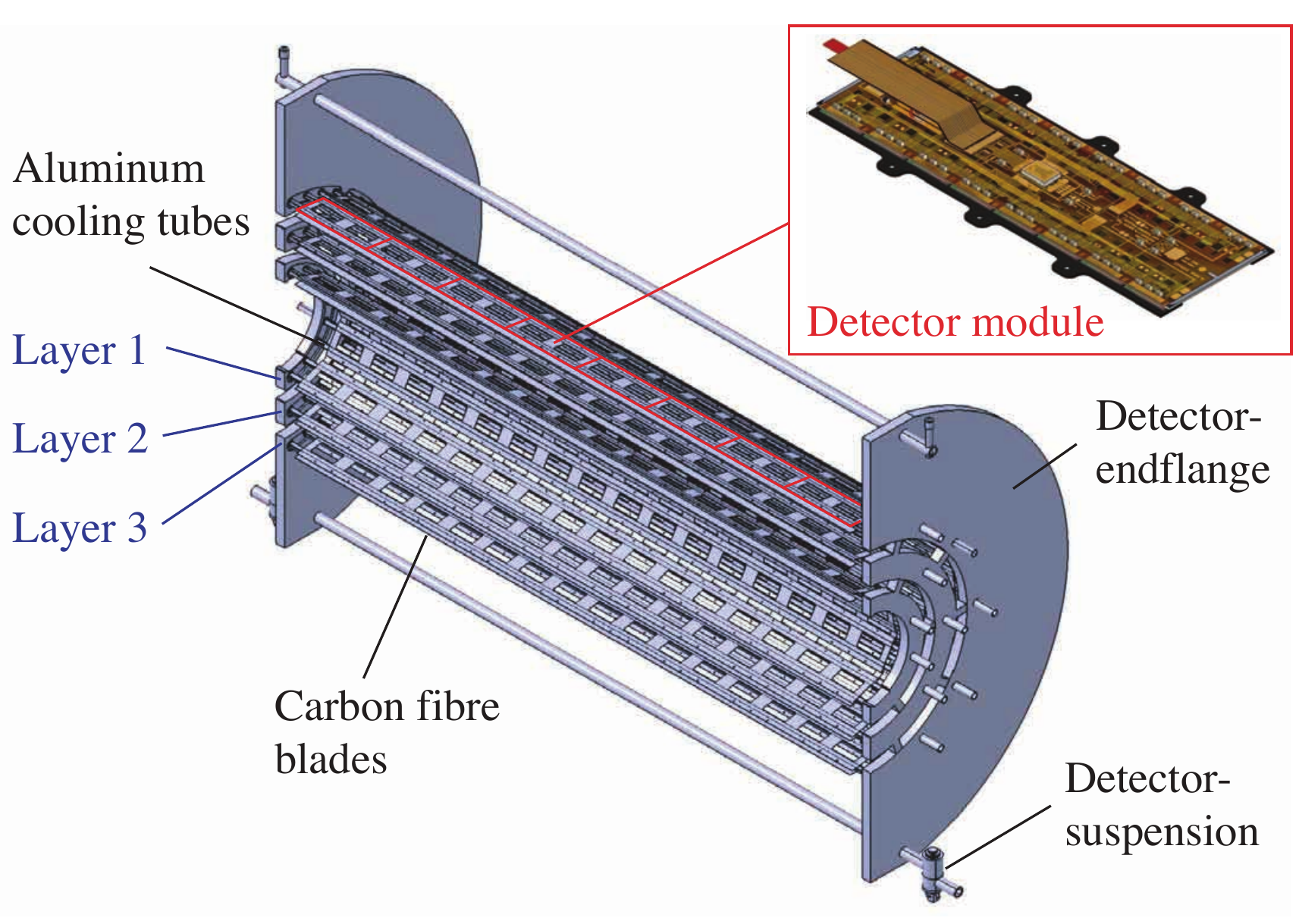}
\caption{Sketch of one half of the pixel barrel detector including the endflanges. \label{fig:halfDetector}}
\end{center}
\end{figure}

The PXB system is constructed in two half-shells  so that  the insertion into the detector after the beam pipe installation is facilitated.  A sketch of one half-shell  is shown in Figure~\ref{fig:halfDetector}. The three half-barrel layers  and the detector endflanges are clearly visible. The arrangement of the detector modules in ladders is indicated. The supply tube is not shown in this picture (see Figure~\ref{fig:SupplyTubeOverview} below).
%%%%%\\\\\\\\\\\\\\\\\\\\\\\\\\\\\\\\\\\\\\\\\\\\
%%%%%%%\\\\\\\\\\\\\\\\\\\\\\\\\\\\\\\\\\\\\\\\\\\\

\begin{figure}[htbp!]
\begin{center}
\includegraphics[width=1.1\textwidth]{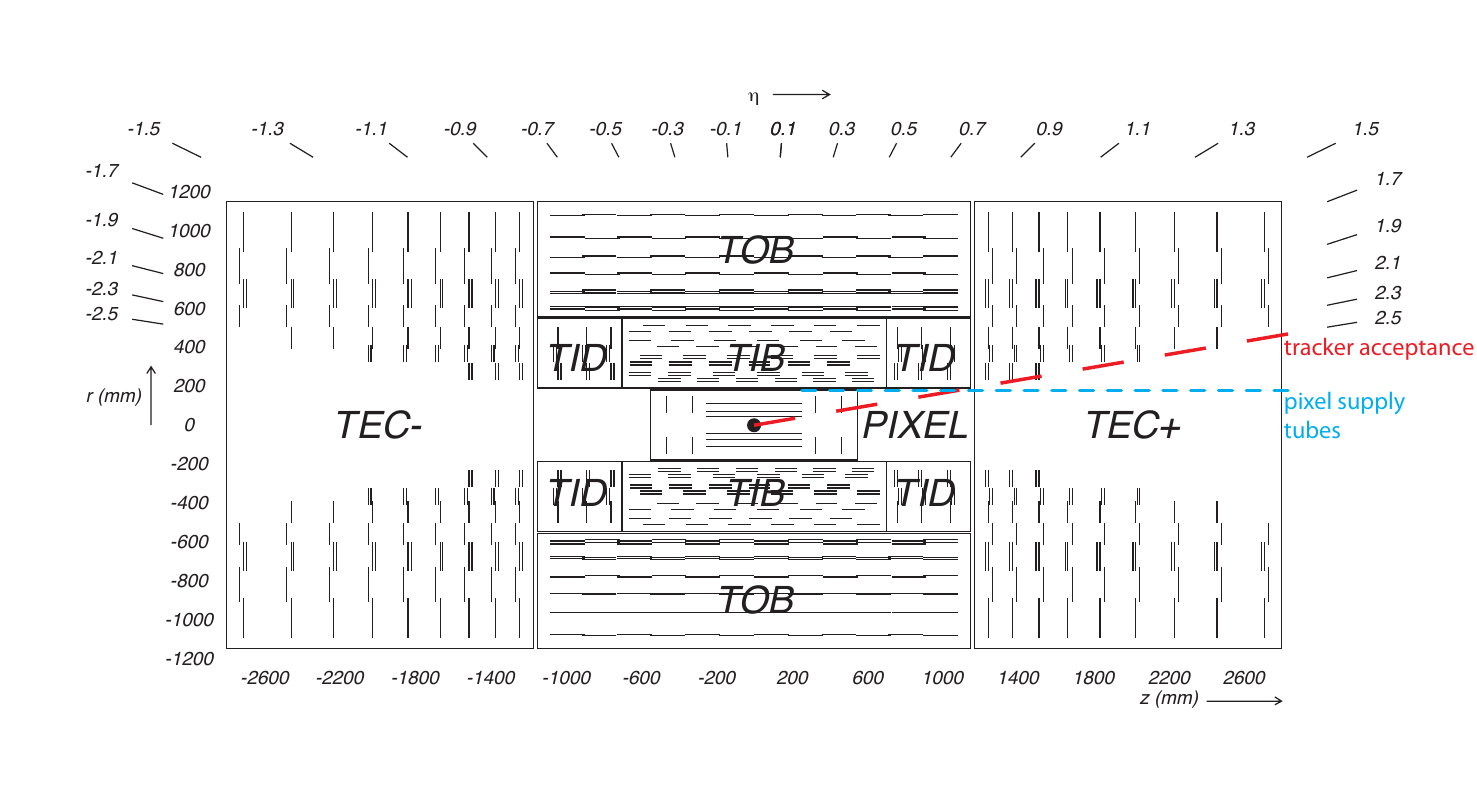}
\caption{Schematic overview of the CMS silicon tracking detectors. The tracker acceptance and the position of the pixel detector supply tubes are illustrated as dashed lines. \label{fig:trackerOverview}}
\end{center}
\end{figure}

A schematic overview of the whole CMS silicon tracking detector is given in Figure~\ref{fig:trackerOverview}. This picture illustrates the dimensions of the pixel detector and its supply tubes with respect to the silicon strip detector.

%%%%%\\\\\\\\\\\\\\\\\\\\\\\\\\\\\\\\\\\\\\\\\\\\
%%%%%%%\\\\\\\\\\\\\\\\\\\\\\\\\\\\\\\\\\\\\\\\\\\\

\subsection{Central Barrel \label{sec:Central}}
\begin{figure}[htbp!]
\begin{center}
\includegraphics[width=0.4\textwidth]{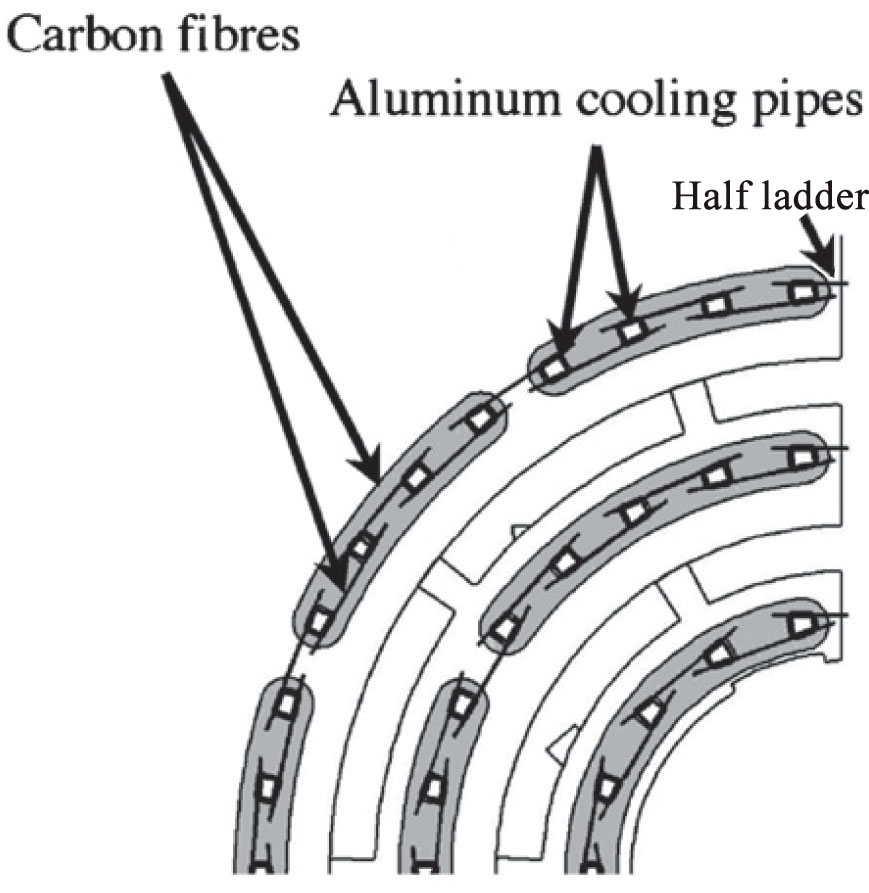}
\caption{Sketch of the carbon fiber ladders mounted on aluminum cooling pipes. \label{fig:barrelLadder}}
\end{center}
\end{figure}

The skeleton of the central barrel is made of aluminum cooling pipes of trapezoidal shape with $\rm 0.3 \ mm$ wall thickness and  $\rm 55.5 \ cm$ length. They are arranged in a way that thin carbon fiber ladders of $\rm 0.22 \ mm$ thickness can be glued on alternating sides as shown in Figure~\ref{fig:barrelLadder} \cite{bib:NIMAoverview}. The detector modules  are then screwed to the carbon fiber ladders. The half-layers 1, 2 and 3 contain 8, 14 and 20  ladders, respectively. The number of cooling pipes is 9, 15 and 21, respectively. Each layer is completed by two half-modules (and half-ladders) at both edges (Figure~\ref{fig:barrelLadder}). 

%The half-size ladders are visible in Figure~\ref{fig:barrelLadder} at the edge of the half-circle.

%%%%%\\\\\\\\\\\\\\\\\\\\\\\\\\\\\\\\\\\\\\\\\\\\
%%%%%%%\\\\\\\\\\\\\\\\\\\\\\\\\\\\\\\\\\\\\\\\\\\\

%\addvspace{1cm}

Eight modules are mounted on each ladder. A module consists of the following components~\cite{bib:NIMA582_776}:
\begin{itemize}
\item $\rm 250 \ \mu m$ thick silicon nitride basestrips supporting the modules,
\item sensors made of $\rm 285 \ \mu m$ thick DOFZ-silicon \cite{bib:SensorDesign},
\item 8 or 16 Read-out Chips (ROCs) \cite{bib:ROCdesign} with $52 \times 80$ pixels of size $150 \times 100 \ \rm\mu m^2$,
\item a High Density Interconnect (HDI), a flexible low mass
three-layer PCB (flex print),
\item a Token Bit Manager (TBM) chip controlling the readout
of the ROCs \cite{bib:NIMA565_73},
\item signal and power cables.
\end{itemize}
A complete module has the dimensions $\rm 66.6 \times 26.0 \ mm^2$ and weighs $\rm 2.2 \ g$ plus up to $\rm 1.3 \ g$ for cables. More than half of the modules weight is due to  the silicon ($\rm 1.4 \ g$) in the sensors and the ROCs. The temperature of the sensors will be maintained at $-10^\circ$C.
Details about the module and sensor design, performance  and  assembly can be found in Ref.~\cite{bib:NIMA582_776,bib:NIMA565_62}.

The signal and power cables run parallel to the modules along the $z$-direction. They are fed through the spacings in the endflange and are then radially distributed until they connect to the Printed Circuit Boards (PCBs) at the detector endflange. 

%Depending on the longitudinal module position, a part of the cables is  attributed to the central barrel region, while the remaining %part belongs to the endflange, as discussed in more detail in Section~\ref{sec:MaterialBudget}.

The envelope of the central barrel region is defined by an internal and an external shielding at 
%The inner- and outermost parts ascribed to the central barrel region are two shieldings at 
$r=3.7 \ \rm cm$ and $r=18.6 \ \rm cm$ extending over the full length of the barrel ($\rm 57 \ cm$). 
Both inner and outer shieldings are made of a $\rm 250 \ \mu m$ thick Kevlar\textsuperscript{\textregistered} cylinder sandwiched between two    $\rm 25 \ \mu m$  aluminum layers.

\subsection{Endflange \label{sec:endflange}}
The endflanges serve several purposes. They act as support structure for the whole barrel, distribute the coolant  and hold electronic devices (endring prints) which are mounted circularly at the outer part of the endflanges. 

\begin{figure}[htbp!]
\begin{center}
\includegraphics[width=0.65\textwidth]{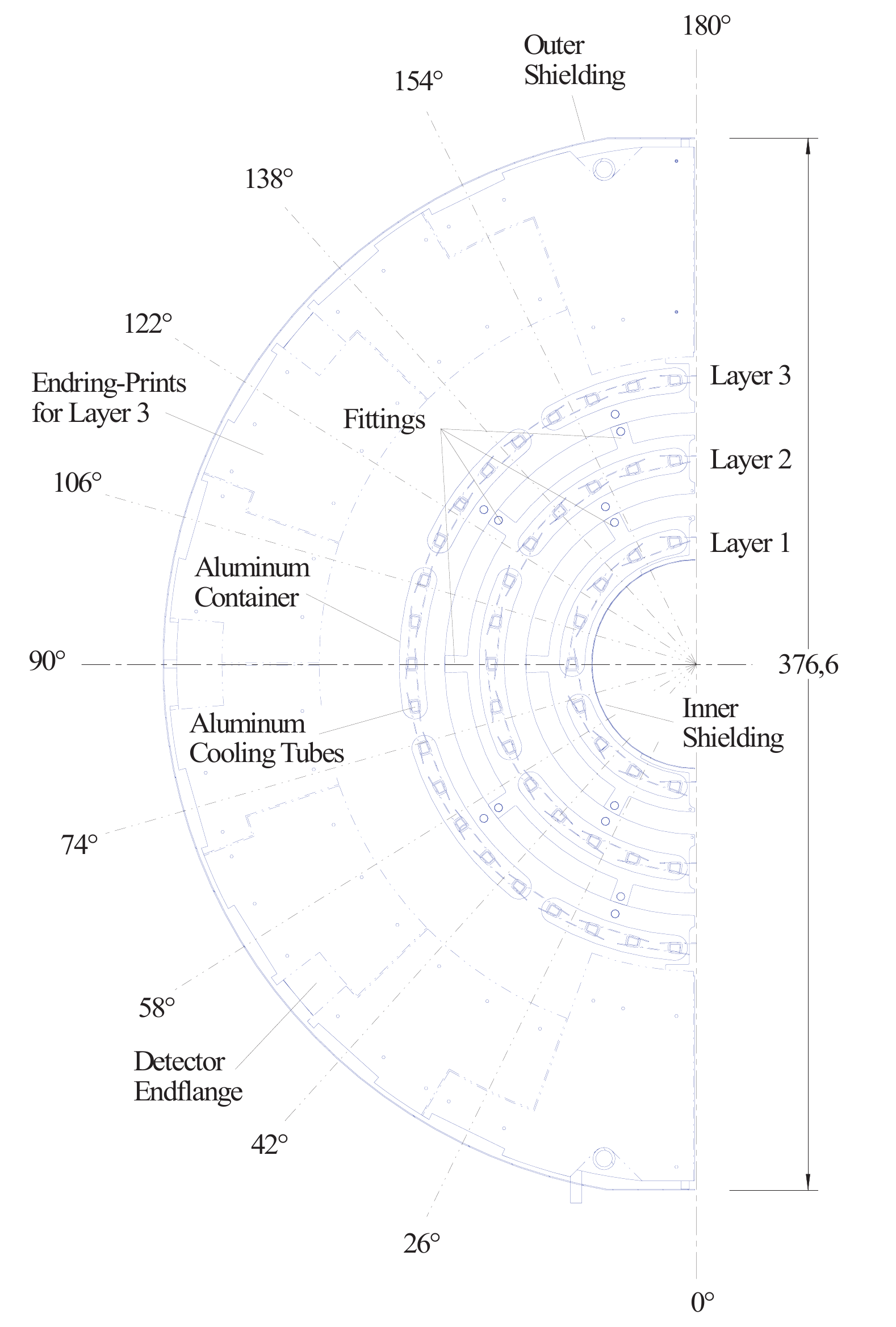}
\caption{Drawing of the layout of the PXB endflange. \label{fig:endflangeDrawing}}
\end{center}
\end{figure}

An illustration of the layout is shown  in Figure~\ref{fig:endflangeDrawing}. Each half-disk contains 10 aluminum containers acting as manifolds  to distribute the cooling fluid. Liquid fluorocarbon ($\rm C_6F_{14}$) is used as coolant in the whole CMS tracker. The trapezoidal aluminum cooling pipes, described in Section~\ref{sec:Central}, are laser welded to the containers with highest precision to guarantee leak tightness of the cooling system \cite{bib:laserWelding}. The side of the endflange facing the supply tube is equipped with ``L'' shaped cooling bends, one for each aluminum manifold,  which facilitate the connection of flexible cooling lines (see Figure~\ref{fig:PXBAndSupplyTube}). 

The half-disks are made of three arcs which are connected by 8 fittings as shown in Figure~\ref{fig:endflangeDrawing}. The space between the arcs is used as feedthrough for the signal cables connecting the modules with the endflange electronics. 

The dimensions of the endflange components are given in Table~\ref{tab:flangeGeometry}. 
%%%%%\\\\\\\\\\\\\\\\\\\\\\\\\\\\\\\\\\\\\\\\\\\\
%%%%%%%\\\\\\\\\\\\\\\\\\\\\\\\\\\\\\\\\\\\\\\\\\\\

%%%%%\\\\\\\\\\\\\\\\\\\\\\\\\\\\\\\\\\\\\\\\\\\\
\begin{table}[htbp]
  \caption{Dimensions of the PXB endflanges in [$\rm cm$]. }
  \label{tab:flangeGeometry}
  \begin{center}
%\begin{small}
  \begin{tabular}{|l|c|c|c|}
   \hline
   	        & Layer 1 & Layer 2 & Layer 3 \\
	\hline 
	inner radius & 3.75  & 6.1  & 9 \\
	outer radius & 5.3   & 8.2  &  18.6  \\
	\hline
	thickness    & \multicolumn{3}{|c|}{0.78} \\ 
	\hline
	  \end{tabular}
%\end{small}
  \end{center}
\end{table}
%%%%%\\\\\\\\\\\\\\\\\\\\\\\\\\\\\\\\\\\\\\\\\\\\

The endflange is made of a thin fiberglass  frame filled with low density Airex\textsuperscript{\textregistered}  foam \cite{bib:Airex}.  Both sides are covered with $\rm 0.5 \ mm$ thick carbon fiber plates,  glued on either side. This technique provides the necessary stability and precision at a very low mass.

The layer 3 arc extends radially up to $\rm 18.6 \ cm$  and holds electronics equipment, the endring prints, which are fixed to the flange with custom made aluminum clamps. They act as interconnect between the modules and the supply tube electronics. The area between the radius $\rm 10.7 \ cm$ and $\rm 18.6 \ cm$ is almost fully covered by the endring PCBs on both sides of the endflange.  They contain low current differential signal (LCDS) transceivers  and cable connectors.  More details about their functionality are available in \cite{bib:NIMA565_73}. 

%%%%%\\\\\\\\\\\\\\\\\\\\\\\\\\\\\\\\\\\\\\\\\\\\
\begin{figure}[htbp]
\begin{center}
\includegraphics[width=0.45\textwidth]{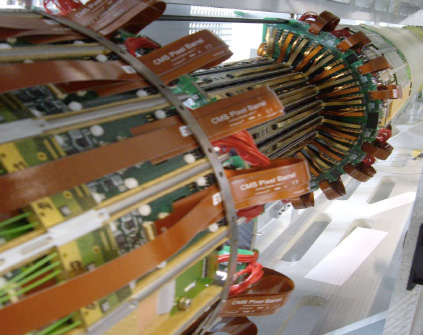}
\caption{Photograph of the PXB detector.  The supply tube is visible on the left half of the picture. The part in the middle shows the third layer of detector modules. The endflange with the radial cabling is visible on the upper right.\label{fig:endflangePhoto}}
\end{center}
\end{figure}
%%%%%%%\\\\\\\\\\\\\\\\\\\\\\\\\\\\\\\\\\\\\\\\\\\\

As visible in Figure~\ref{fig:endflangePhoto}, the cables cover a large fraction of the endring prints and contribute a significant part to the total amount of metal in the vicinity of the endflange. This is discussed in more detail in Section~\ref{sec:MaterialBudget}.

All connections between the barrel detector and the supply tube  run through the barrel endflange. The connections are flexible to a certain extent to allow small movements of the pixel detector. Also the cooling pipes connecting to the manifolds in the endflange  are made of  flexible material, silicone rubber   reinforced with fiberglass. These cooling lines are distributed radially in the $r-\phi$ plane at $z=\pm 30.5 \ \rm cm$ and are fixed with steel clamps to the ``L'' shaped cooling bends on the endflange and supply tube.

\subsection{Supply Tube \label{sec:supplyTube}}
\begin{figure*}[bp]
\begin{center}
\includegraphics[width=0.7\textwidth]{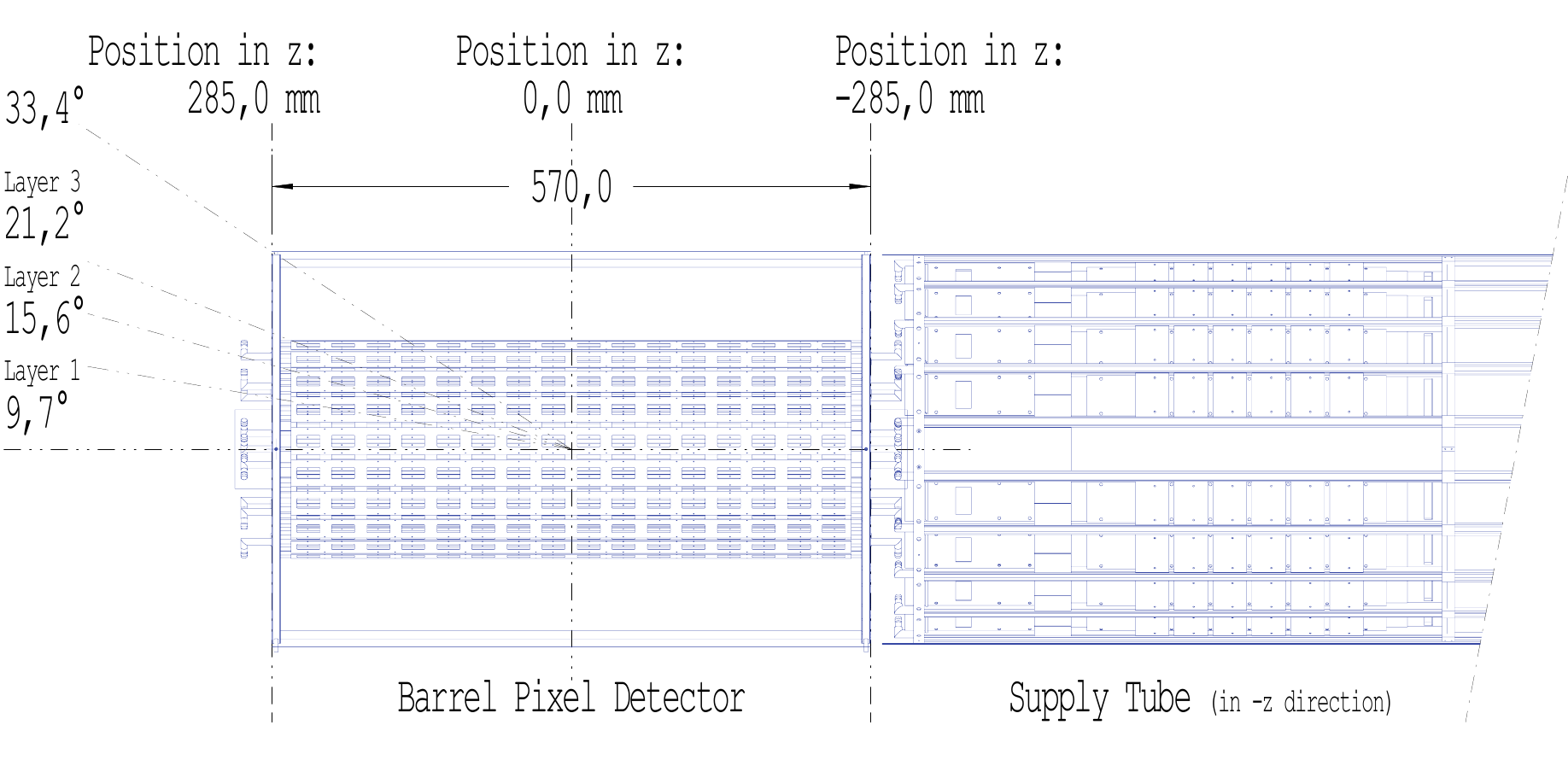}
\caption{Pixel barrel detector (on the left) connecting to the supply tube (on the right). The "L" shaped cooling bends are visible just outside the rectangle on the left edge, and between barrel and supply tube.\label{fig:PXBAndSupplyTube}}
\end{center}
\end{figure*}

\begin{figure*}[htbp]
\begin{center}
\includegraphics[width=0.9\textwidth]{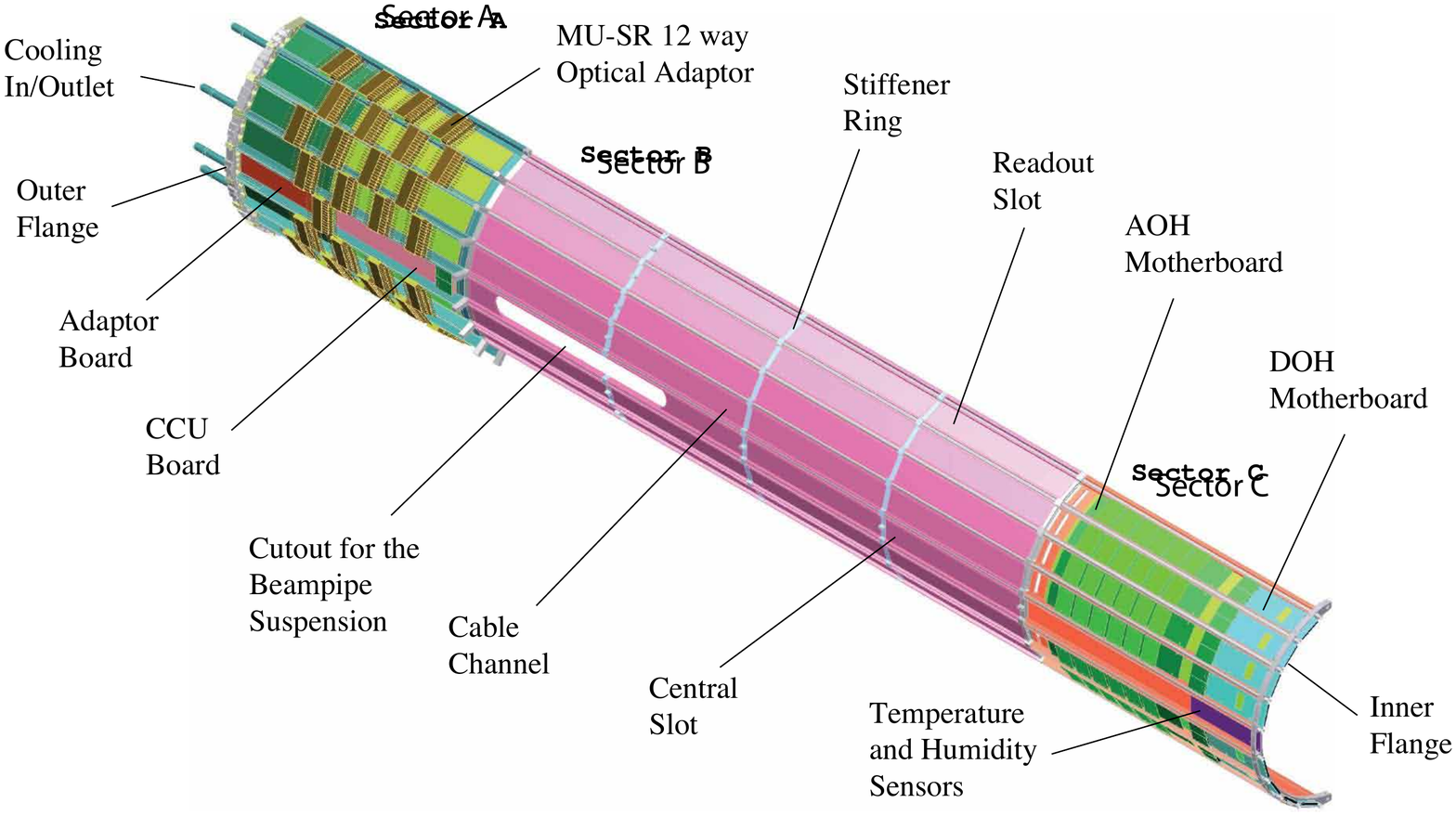}
\caption{Illustration of the pixel barrel supply tube half-shell. It is subdivided into three logical sectors. Sector A is the leftmost part. The part in the middle (the four segments are visible) is Sector B. The innermost part (connecting to the endflange) is labelled Sector C.\label{fig:SupplyTubeOverview}}
\end{center}
\end{figure*}

The supply tube, one at each end of the pixel barrel, connects  the PXB detector to the outside world. It contains the power cables, optical signal cables, and cooling lines. The tube is constructed in two half shells at both ends of the detector.   Its position with respect to the barrel is shown in Figure~\ref{fig:PXBAndSupplyTube}. The tube has a length of $\rm 2.2 \ m$ and an inner radius of $\rm 18 \ cm$ with a wall thickness between $\rm 1 \ cm$ and $\rm 3 \ cm$, depending on the $z$-position. 
%%%%%\\\\\\\\\\\\\\\\\\\\\\\\\\\\\\\\\\\\\\\\\\\\
%%%%%%%\\\\\\\\\\\\\\\\\\\\\\\\\\\\\\\\\\\\\\\\\\\\
%%%%%%%%%%
It is subdivided into three logical sectors (sectors A, B and C),  as shown in Figure~\ref{fig:SupplyTubeOverview}. The innermost part, Sector C, as well as a small fraction of Sector B, are inside of the tracker acceptance as illustrated in Figure~\ref{fig:trackerOverview}.

%%%%%\\\\\\\\\\\\\\\\\\\\\\\\\\\\\\\\\\\\\\\\\\\\
%%%%%%%%%%
%%%%%%%\\\\\\\\\\\\\\\\\\\\\\\\\\\\\\\\\\\\\\\\\\\\

The supply tube also hosts several electronic devices, which are briefly described in the following:
for the readout, the Analog Optical Hybrid (AOH) converts
the analog signals to optical signals. The AOH includes
 the Analog Level Translator (ALT) chips.
Six AOHs are placed in a single analog mother board.

The signals controlling the detector are converted from optical to digital (electrical) signals
in the Digital Optical Hybrid (DOH).
The Phase Locked Loop (PLL) chips are used to split the clock
from the trigger and the
DELAY25 chips adjust the relative phases of all control signals.
The Gate-Keeper chips block the idle patterns transmitted to keep the links balanced and they
 convert the Low Voltage Differential Signals (LVDS)  used by PLLs and DELAY25s
to low current differential signals (LCDS) used by the pixel
front-end chips. 
Two DOHs, 2 PLLs, 2 DELAY25s and 2 Gate-Keepers are placed on
a single digital-motherboard.

The analog and digital mother boards are plugged into a single
board called the opto-board.
%There is one opto-board per supply tube sector.

To configure the AOHs and DOHs, the CCU chip 
(Communication and Control Unit) is used. For each half-shell of the supply
tube a CCU-board, with 9 CCUs and 9 LVDS-MUX chips is used.
More details can be found in \cite{bib:NIMA565_73}. 

The supporting elements of the  structure are the stainless steel cooling tubes with a wall thickness of 0.1~mm running along the $z$-direction. They are connected to the stiffener rings (fiberglass) and the inner and outer  aluminum flanges. The gaps in-between are filled with foamed material (Airex\textsuperscript{\textregistered})  to guarantee the required rigidity. All power and slow control lines are embedded in the supply tube body. This allows a clear layout of the wiring and makes the system also more reliable.

\section{Analysis of the Detector Material \label{sec:MaterialBudget}}
The amount of material given in the following tables was partially obtained by direct weight measurements of the  components and partially calculated from design drawings. For instance, only the final total weight of the endflange support structure (without aluminum manifolds, electronics and coolant) could be determined by measurement due to the complexity of the construction process. Another example is the fraction of copper in the printed circuit boards which needed to be estimated based on the design drawings. Therefore the detailed composition of the various material fractions is subject to uncertainties  in several cases. The calculated total weights of the substructures  agree  within a few percent with the real weights of the detector components. The obtained precision is therefore sufficient for an adequate description of the detector in the Monte Carlo simulation.

The following tables summarize the amount of material in all of the detector components, according to the order of discussion in the previous sections.

The material budget for the central barrel support structure is summarized in Table~\ref{tab:BarrelMaterial}, which shows the total amount of aluminum, carbon fiber and coolant. The total material budget of the modules is given in Table~\ref{tab:ModuleMaterial}.

%%%%%\\\\\\\\\\\\\\\\\\\\\\\\\\\\\\\\\\\\\\\\\\\\
\begin{table}[h!]
  \caption{Amount of material contained in the central barrel support structure (without modules and endflange)  per half-shell in units of [$\rm g$].}
  \label{tab:BarrelMaterial}
  \begin{center}
%\begin{small}
  \begin{tabular}{|l|c|c|c|}
   \hline
   	        & Layer 1 & Layer 2 & Layer 3 \\
	\hline 
	carbon fiber & 40   &  65   &  91   \\
	aluminum    &  52  &  86   & 120    \\
	coolant      &  75  &  126  & 176   \\
	   \hline
  \end{tabular}
%\end{small}
  \end{center}
\end{table}
%%%%%\\\\\\\\\\\\\\\\\\\\\\\\\\\\\\\\\\\\\\\\\\\\

%%%%%\\\\\\\\\\\\\\\\\\\\\\\\\\\\\\\\\\\\\\\\\\\\
\begin{table}[h!]
  \caption{Total amount of material contained in the detector modules in [$\rm g$] per detector half-shell. The sensors and ROCs are made of silicon. The module baseplates are made of silicon nitride and steel for screws and nuts. The materials Kapton\textsuperscript{\textregistered}, copper, nickel, gold and solder belong to the HDI and its electrical components. Barium titanate is contained in the capacitors. }
  \label{tab:ModuleMaterial}
  \begin{center}
%\begin{small}
  \begin{tabular}{|l|c|c|c|}
   \hline
   	        & Layer 1 & Layer 2 & Layer 3 \\
	\hline 
	silicon  & 100  & 166    & 232    \\
    silicon nitride  & 43  & 67    &  90   \\
    steel &            7  &  12  &   17   \\
 %   silicone gel  &  2.5   &  3.8  & 5.1   \\
	 glue        &  2.3    &  3.6   &  4.8  \\ 
 Kapton\textsuperscript{\textregistered}  & 5.6   &  9.2 	&  12.9  \\
  copper  & 4.1   &  6.8   & 9.6  \\
  nickel &  0.9   &  1.4   & 2   \\
 gold   & 0.4   & 0.7   &  1  \\
 solder & 4.8    &  7.7   &  10.6  \\
 barium titanate  & 12   &  19.7   &  27.4   \\

	   \hline
  \end{tabular}
%\end{small}
  \end{center}
\end{table}
%%%%%\\\\\\\\\\\\\\\\\\\\\\\\\\\\\\\\\\\\\\\\\\\\

%%%%%%%%%%%%%%\\\\\\\\\\\\\\\\\\\\\\\\\\\\
\begin{table}[h!]
  \caption{Amount of material in signal and power cables of the central barrel region in [$\rm g$] per detector half-shell.}
  \label{tab:signalCable}
  \begin{center}
%\begin{small}
  \begin{tabular}{|l|c|c|c|}
   \hline
   	        & Layer 1 & Layer 2 & Layer 3 \\
	\hline 
 Kapton\textsuperscript{\textregistered}  & 15   &  24 	&  34  \\
  copper  & 12   &  19   & 27  \\
  aluminum & 7  & 11  & 15  \\
	   \hline
  \end{tabular}
%\end{small}
  \end{center}
\end{table}
%%%%%%%%%%%%%%%%%%%%%%%%%%%%%%%%%%%%%////////////

%\addvspace{1cm}

Since a ladder holds eight modules, there are four different cable lengths in the central barrel region: 23, 16, 9 and 3 $\rm cm$. The power cables consist of an aluminum core with a copper and Kapton\textsuperscript{\textregistered} coating, while the flat signal  cables are made of 22 copper traces and a ground mesh in a Kapton\textsuperscript{\textregistered} coating. Together, the cables have a mass occupancy of $\rm 0.034 \ g/cm$. Since the total cable length in the barrel is about $\rm 50 \ m$ per half-shell, they contribute to around $\rm 170 \ g$ to the material budget. The decomposition of the various materials for these cables is given in Table~\ref{tab:signalCable}.

The inner shielding of the PXB detector weighs $\rm 65 \ g$ ($\rm 17 \ g$ aluminum and $\rm 48 \ g$ Kevlar\textsuperscript{\textregistered}), while the outer shield weighs $\rm 320 \ g$ ($\rm 87 \ g$ aluminum and $\rm 233 \ g$ Kevlar\textsuperscript{\textregistered}).

The endring PCBs mounted on the detector endflange (Table~\ref{tab:flangeMaterial}) consist of standard epoxy substrate in five layers of $\rm 68 \ \mu m$ thickness and two (four) copper layers of $\rm 18 \ \mu m$ ($\rm 25 \ \mu m$) thickness. Their material budget is given in the upper rows of Table~\ref{tab:endringPrintsMaterial}.

As mentioned earlier, the signal cables connecting the modules with the endring electronics are running  parallel to the beam axis above and below the modules until the cables reach the endflange (this material is given in Table~\ref{tab:signalCable}). Depending on the longitudinal module position, a part of the cables is  attributed to the central barrel region, while the remaining part belongs to the endflange.  The  length fractions on the endflange depend on the radial position of the corresponding connectors and the position of the feedthroughs. The material fraction which is attributed to the endflange is given in the middle part of Table~\ref{tab:endringPrintsMaterial}.

A similar type of Kapton\textsuperscript{\textregistered} cables, but with different width and length is used to connect the endring prints with the supply tube. In addition, power cables made of copper and aluminum, are connected at the edge of the endflange.
These are summarized in the bottom part of Table~\ref{tab:endringPrintsMaterial}.

The enflange and supply tube are also connected by cooling pipes.
 The pipes are made of silicone rubber ($\rm 46 \ g$ per half-disk)  reinforced with fiberglass, containing $\rm 53 \ g$ of coolant.

The material decomposition for the three sectors of the supply tube is shown in Table~\ref{tab:materialSupTub}. The numbers for the supply tube  are obtained by evaluation of the relative material fractions based on the true weight of the final structure. This means that, by definition, the total weight of the supply tube in the simulation is in perfect agreement with the true weight, while only the material fractions are subject to uncertainties.
In contrast to this, the numbers given in Tables~\ref{tab:BarrelMaterial} to \ref{tab:endringPrintsMaterial} are absolute values obtained by direct weight measurements or calculations based on the design drawings.  These different approaches are motivated by the different granularity of the implementations in the software as discussed in the following section.

%%%%%\\\\\\\\\\\\\\\\\\\\\\\\\\\\\\\\\\\\\\\\\\\\
\begin{table}[h]
  \caption{Material decomposition of the PXB endflange in [$\rm g$] per half-disk. }
  \label{tab:flangeMaterial}
  \begin{center}
%\begin{small}
  \begin{tabular}{|l|c|c|c|}
   \hline
   	        & Layer 1 & Layer 2 & Layer 3 \\
	\hline 
	carbon fiber & 4  &  8.5   &  75   \\
	glass fiber &  5.6  &  9   &  17  \\
	Airex\textsuperscript{\textregistered}        & 0.4  &  1.3     & 21     \\
	aluminum    &  5.3 &  8.4   &  12   \\
	coolant       &  8   &   15.5   &   17.5   \\
	 steel        & 5  &   7.5   &    12.5  \\
	   \hline
  \end{tabular}
%\end{small}
  \end{center}
\end{table}
%%%%%\\\\\\\\\\\\\\\\\\\\\\\\\\\\\\\\\\\\\\\\\\\\

%%%%%\\\\\\\\\\\\\\\\\\\\\\\\\\\\\\\\\\\\\\\\\\\\
\begin{table}[h]
  \caption{Material decomposition of the endring electronics including the cabling going to supply tube and modules in [$\rm g$] per half-disk. The amount of solder is estimated. The term ``plastics'' refers to various types of polymers of similar density which are used in connectors and cable insulations. }
  \label{tab:endringPrintsMaterial}
  \begin{center}
%\begin{small}
  \begin{tabular}{|l|c|c|}
   \hline
         & boards layer 1+2 & boards layer 3\\
        \hline
   	   copper     & 32  & 27  \\
	   	FR4        & 26   & 22  \\
		plastics     & 33  & 15    \\
		solder     &  2.6   &  2.2    \\
	    \hline
	     & \multicolumn{2}{|c|}{cables to modules} \\
	     & signal cables & power cables \\
	     \hline
	     copper & 12.1    &  9.2   \\
	     Kapton\textsuperscript{\textregistered} &  26   & 1.25    \\
	     aluminum & -  &  11.7   \\
          \hline
	 	  & \multicolumn{2}{|c|}{cables to supply tube}\\
	      & power cables& signal cables \\
	      \hline
	     copper & 29  & 24    \\
	     Kapton\textsuperscript{\textregistered} & - & 28   \\
	     aluminum & 16  &  - \\
	     plastics  & 55   & 8  \\
	   	     \hline
	     \end{tabular}
%\end{small}
  \end{center}
\end{table}
%%%%%\\\\\\\\\\\\\\\\\\\\\\\\\\\\\\\\\\\\\\\\\\\\

%%%%%\\\\\\\\\\\\\\\\\\\\\\\\\\\\\\\\\\\\\\\\\\\\
\begin{table}[h]
  \caption{Material decomposition of the three sectors of the supply tube in [$\rm g$] per half-shell.  Only Sector C and a small fraction of Sector B are within the tracker acceptance. The term ``plastics'' refers to different kinds of polymers of similar density such as polyesther, polyvinylchloride and polyoxymethylene.}
  \label{tab:materialSupTub}
  \begin{center}
%\begin{small}
  \begin{tabular}{|l|c|c|c|}
   \hline
   	   sector     & A & B & C  \\
	   \hline
	   aluminum & 610  &  746    & 470    \\
	  FR4        & 653  & 517    &  517    \\
	  Airex\textsuperscript{\textregistered}       &  80  & 69    &   30    \\
	  steel       &  351  & 223   &  94    \\
	  copper      &  51  & 283  &  150     \\
	  plastics    &  610  & 1740    &  720   \\
	  coolant     &   330  &    828  & 346  \\  
			   \hline
  \end{tabular}
%\end{small}
  \end{center}
\end{table}
%%%%%\\\\\\\\\\\\\\\\\\\\\\\\\\\\\\\\\\\\\\\\\\\\

\section{Monte Carlo Simulation\label{sec:MCsimulation}}
In this section we describe the implementation of the material budget in the simulation software.

In the CMS software framework,  the passage of particles through matter is simulated by the GEANT4 toolkit \cite{bib:GEANT4A,bib:GEANT4B}. Therefore, the exact position, amount and type of materials need to be specified. The access to the detector components and their geometry is provided by the Detector Description Database (DDD) \cite{bib:DDD}. The master data source for the DDD is a collection of files implemented in the Extensible Markup Language (XML) using the Detector Description Language. The  detector is divided into logical volumes of one particular type of material. In many cases this  is a mixture of the materials contained in the defined volume. For instance, screws are usually not represented as single volumes, but their material is mixed into the definition of a larger structure containing them.

The description in the Monte Carlo simulation needs to be as precise as possible in order to reproduce the exact amount of material interactions. Such interactions deteriorate the trajectories of particles and influence the measurements in the outer detectors (strip tracker, calorimeters and muon detectors). The  sensitive volumes in the barrel have been described with the highest possible accuracy (Section~\ref{sec:MCBarrel}). The material closer to the interaction region is more relevant than the material further away, therefore, the high level of detail in the central barrel is driven by the innermost layer. Also the endflanges are described with a high level of detail, since they are located directly in front of the sensitive volumes of the pixel forward detector. This is discussed in Section~\ref{sec:MCendflange}. 

The supply tubes extend to the very forward region and cover a pseudo-rapidity of $1.2 < |\eta| < 3.4$.   The total material in the tracker reaches its maximum at a pesudo-rapidity of $|\eta| = 1.2$ as discussed in Section~\ref{sec:result}. Hence, small variations in this region have a smaller impact on the total material budget than in the central barrel. Therefore, the supply tubes are described in less detail  (Section~\ref{sec:MCsupply}).

In the following, the implementation of the detector geometry is outlined. The software representation is based on the material content discussed in the previous sections.

\subsection{Central Barrel\label{sec:MCBarrel}}
\begin{figure}[h]
\begin{center}
\includegraphics[width=0.65\textwidth]{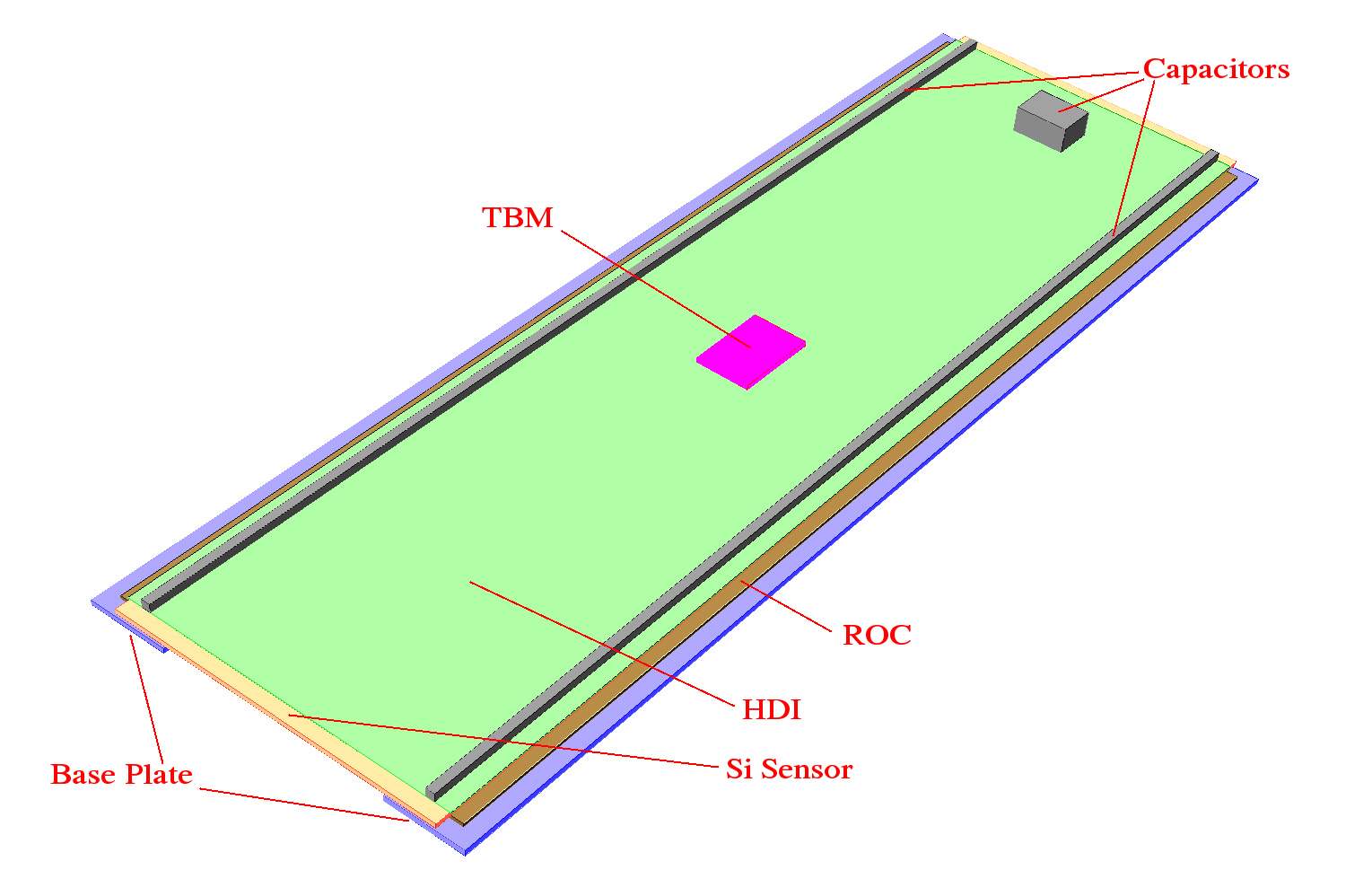}
\caption{A full barrel pixel module in the Monte Carlo simulation. 
%``TBM'' is the Token Bit Manager chip. ``HDI'' refers to the High Density Interconnect. 
 \label{fig:pixBarMCmodule}}
\end{center}
\end{figure}

A pixel barrel module is represented by
several boxes of different thicknesses  stacked on top of each other. Their dimensions are: 
\begin{itemize}
\item base plates: $\rm 0.45 \ cm \times 6.6 \ cm \times  300 \ \mu m$, There are two baseplates per full module positioned at the module edges, 
\item silicon sensors: $\rm 1.9 \ cm \times 6.6 \ cm \ \times 285 \ \mu m$,
\item High Density Interconnect (HDI): $\rm 1.9 \ cm \times 6.5 \ cm \times 44 \ \mu m$,
\item Read-Out Chip (ROC): $\rm 2 \ cm \times 6.5 \ cm \times 180 \ \mu m$.
\end{itemize}
Four additional cuboidal volumes are defined for each module to 
describe  localized components such as capacitors and TBM chips.
These are visualized in Figure~\ref{fig:pixBarMCmodule}. The capacitors along the module edge are implemented as  strips of $\rm 6.5 \ cm$ length and $\rm 0.5 \ mm \times 0.5 \ mm$ cross section.
%%%%%\\\\\\\\\\\\\\\\\\\\\\\\\\\\\\\\\\\\\\\\\\\\
%%%%%%%\\\\\\\\\\\\\\\\\\\\\\\\\\\\\\\\\\\\\\\\\\\\

The modules are replicated  to form ladders and
then layers, starting from the single module description contained in
the XML.

The  cooling pipes of the central barrel support structure have trapezoidal cross sections  with various inclination angles (see Figure~\ref{fig:barrelLadder}). The angles are computed automatically by the software given the number of ladders and the corresponding radius of the layer. The carbon fiber ladders holding the modules are represented as rectangles of $\rm 2.6 \ cm$ width and $\rm 250 \ \mu m$ thickness, running along the full length of the barrel ($\rm 53 \ cm$). There are 16 holes in the ladders, two under each module, of $\rm 2.2 \ cm$ length and $\rm 1.1 \ cm$ width to reduce the amount of material.

For the cables,
we use four different types of volumes 
to account for the different lengths ($23$, $16$, $9$ and $\rm 3 \ cm$),  reaching modules at different longitudinal positions. Cable volumes
are boxes with  cross sections of $\rm 300 \ \mu m \times 0.67 \ cm$  running from
the module (at a distance of $\rm 1.93 \ cm$ from the module edge) to the endflange. The remaining
length of the cable is accounted for in the endflange description. The volumes of the cables  are
replicated  in their proper positions to form the layers.

The  half-ladders are described in a similar way and their
volumes are located in the top and bottom edges of each half-shell.

\subsection{Endflange\label{sec:MCendflange}}
\begin{figure}[h]
\begin{center}
\includegraphics[width=0.58\textwidth]{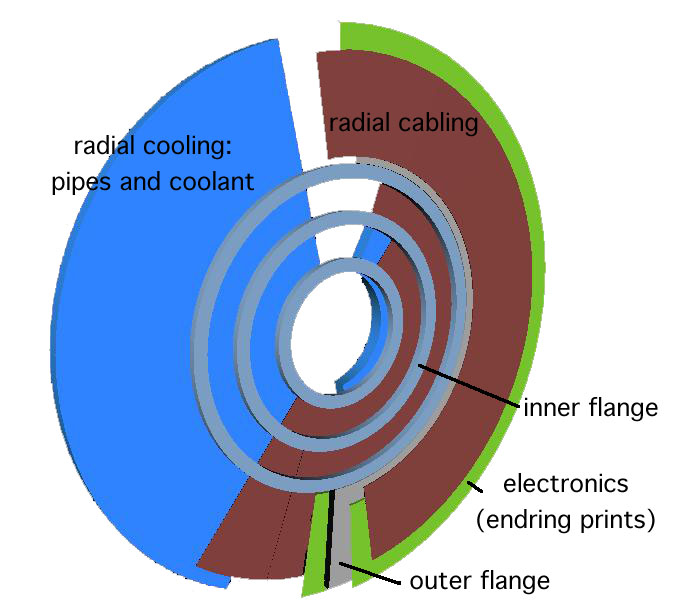}
\caption{Cross section of the endflange in the Monte Carlo simulation. The various volumes are clipped for better visibility (except for the radial cooling). \label{fig:endflangeMC}}
\end{center}
\end{figure}

The logical volumes forming the endflange are implemented as circular disks and rings, except for the cooling, which only covers an azimuthal angle of $\Delta\phi = 140^\circ$ (Figure~\ref{fig:endflangeMC}).  They are located at a distance $z = \pm 28 \ \rm cm$ from the detector center.  

The parts labelled ``inner flange'' represent the arcs belonging to layers 1 and 2. Their material  is composed of the  glass fiber and carbon fiber frame, including the aluminum manifolds, and the contained coolant. The innermost part of the outer flange is implemented  analogously  (third ring in Figure~\ref{fig:endflangeMC}). The second part of the outer flange extends further in radius (up to $\rm 18.6 \ cm$) and does not contain any cooling manifold. 

The endring prints and cables are mounted onto the outer flange. The cables between the detector modules and the endring prints are labelled ``radial cabling''. The radial cabling extends up to a maximal radius of $\rm 17 \ cm$ as visible in Figure~\ref{fig:endflangePhoto}. This volume contains the remaining fraction of the central barrel signal cables.

There is an additional volume (not shown in Figure~\ref{fig:endflangeMC}) implemented as a ring at $z= \pm 30.5 \ \rm cm$ and $r = 18 \ \rm cm$ which contains the signal and power cables connecting the endring prints with the supply tube. This ring also includes the innermost part  of the supply tube which is labelled ``Inner Flange'' in Figure~\ref{fig:SupplyTubeOverview}.

%%%%%\\\\\\\\\\\\\\\\\\\\\\\\\\\\\\\\\\\\\\\\\\\\
%%%%%%%\\\\\\\\\\\\\\\\\\\\\\\\\\\\\\\\\\\\\\\\\\\\

\subsection{Supply Tube\label{sec:MCsupply}}
The representation of the supply tube is straightforward. Each of the three sectors described in Section~\ref{sec:supplyTube} and Figure~\ref{fig:SupplyTubeOverview} is represented as a single volume with an average material content according to Table~\ref{tab:materialSupTub}. There are three cylinders, one for each sector  with inner radius of $r=18 \ \rm cm$ and wall thickness of $2.5 \ \rm cm$. The lengths of Sectors A, B and C are  $\rm 48.5$, $\rm 120$ and $49.3 \ \rm cm$, respectively.

As discussed in the last paragraph of the previous section, the circular ``Inner Flange'' of the supply tube is implemented as a separate ring which also  contains the steel connections for the cooling pipes and cabling.

\section{Results and Summary\label{sec:result}}
\begin{figure}[h]
\begin{center}
\includegraphics[width=0.58\textwidth]{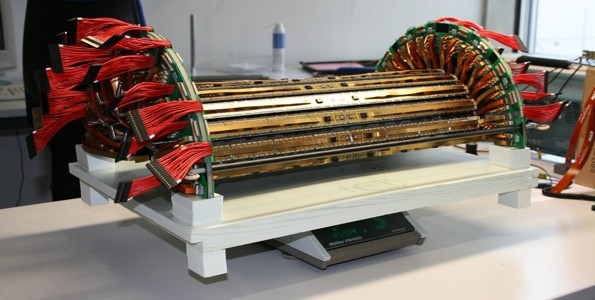}
\caption{The completed PXB half-shell on a balance which shows the weight of $\rm 2598 \ g$. Not included are the coolant and the signal cables  to the supply tube. \label{fig:balance}}
\end{center}
\end{figure}

After the final assembly of the pixel barrel detector, the weight of one completed half-shell was measured (without coolant). Figure~\ref{fig:balance} shows one final half-detector on a balance.
The cables and cooling tubes between supply tube and detector endflange were not included in this measurement. The result was $\rm 2598 \ g \ \pm \ 0.5 \ g$.

For comparison, the weight of this configuration in the Monte Carlo simulation is $\rm 2455 \ g$. The remaining disagreement is around 6\% and is due to the unavoidable approximations  as discussed in Section~\ref{sec:MaterialBudget}.

\begin{figure*}[p]
\begin{center}
\includegraphics[width=1.03\textwidth]{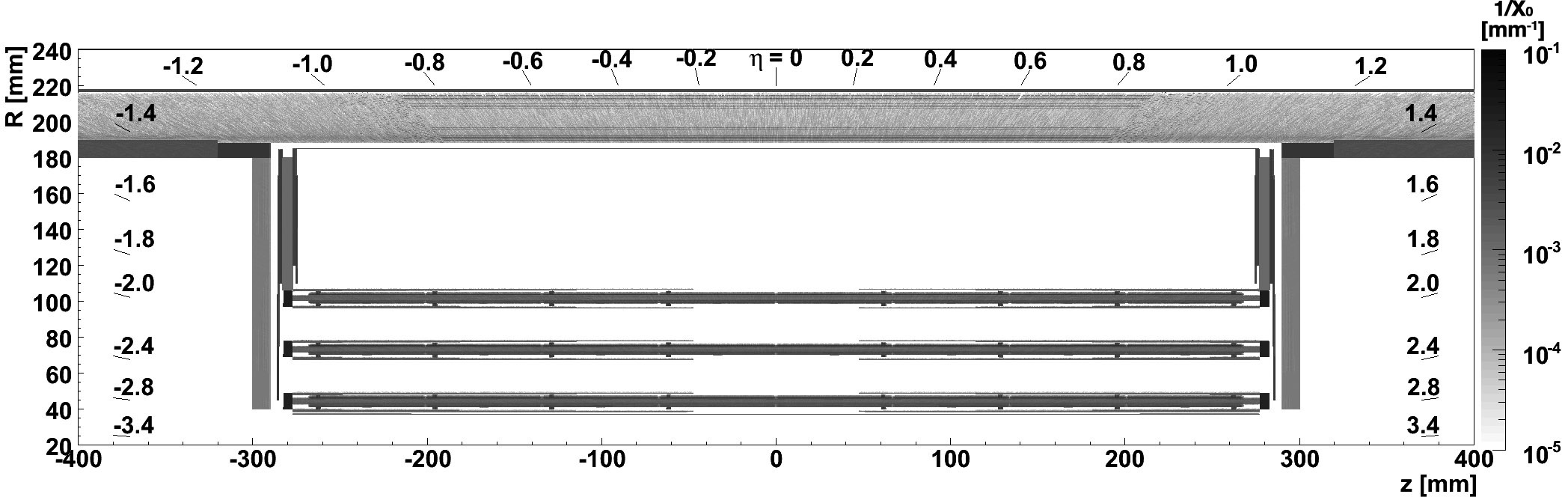}
\caption{$r-z$ profile of the PXB  detector (as implemented in the simulation). The grayscale indicates the inverse of the radiation length $1/X0(r,z,\phi)$ averaged over $\phi$, at a given point in space. 
The pseudo-rapidity $\eta$ is visualized near the upper, left and right edges of the diagram. \label{fig:RadLengthProfile}}
\end{center}
\end{figure*}

Figure~\ref{fig:RadLengthProfile} shows a $r-z$ profile of the detector as it is implemented in the Monte Carlo simulation. In this and the following plots, the material is averaged over the azimuthal angle $\phi$.

\begin{figure*}[p]
\begin{center}
\includegraphics[width=1.03\textwidth]{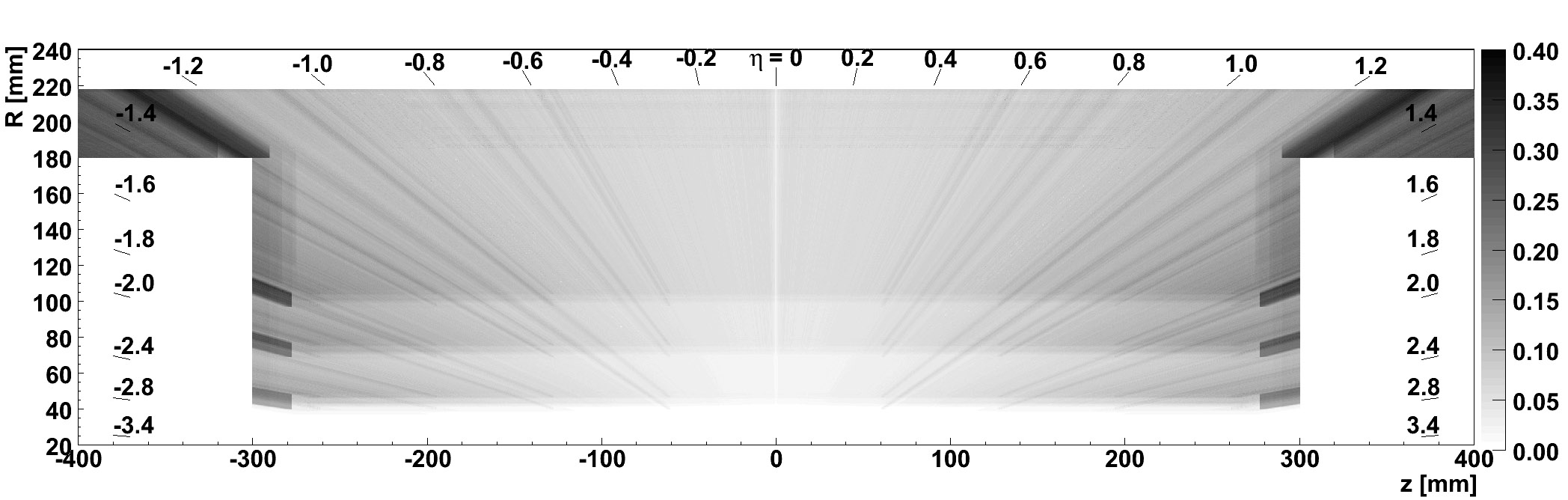}
\caption{Integrated radiation length $x/X_0$ experienced by particles emerging from the interaction point as a function of pseudo-rapidity, averaged along $\phi$. The grayscale indicates the number of  radiation lengths traversed by a particle propagating on a straight line.  \label{fig:RadLengthProfileSum}}
\end{center}
\end{figure*}

The integrated material as experienced by  particles emerging from the interaction point and traveling on straight lines through the detector is shown in Figure~\ref{fig:RadLengthProfileSum}.

\begin{figure*}[p]
\begin{center}
\includegraphics[width=0.48\textwidth]{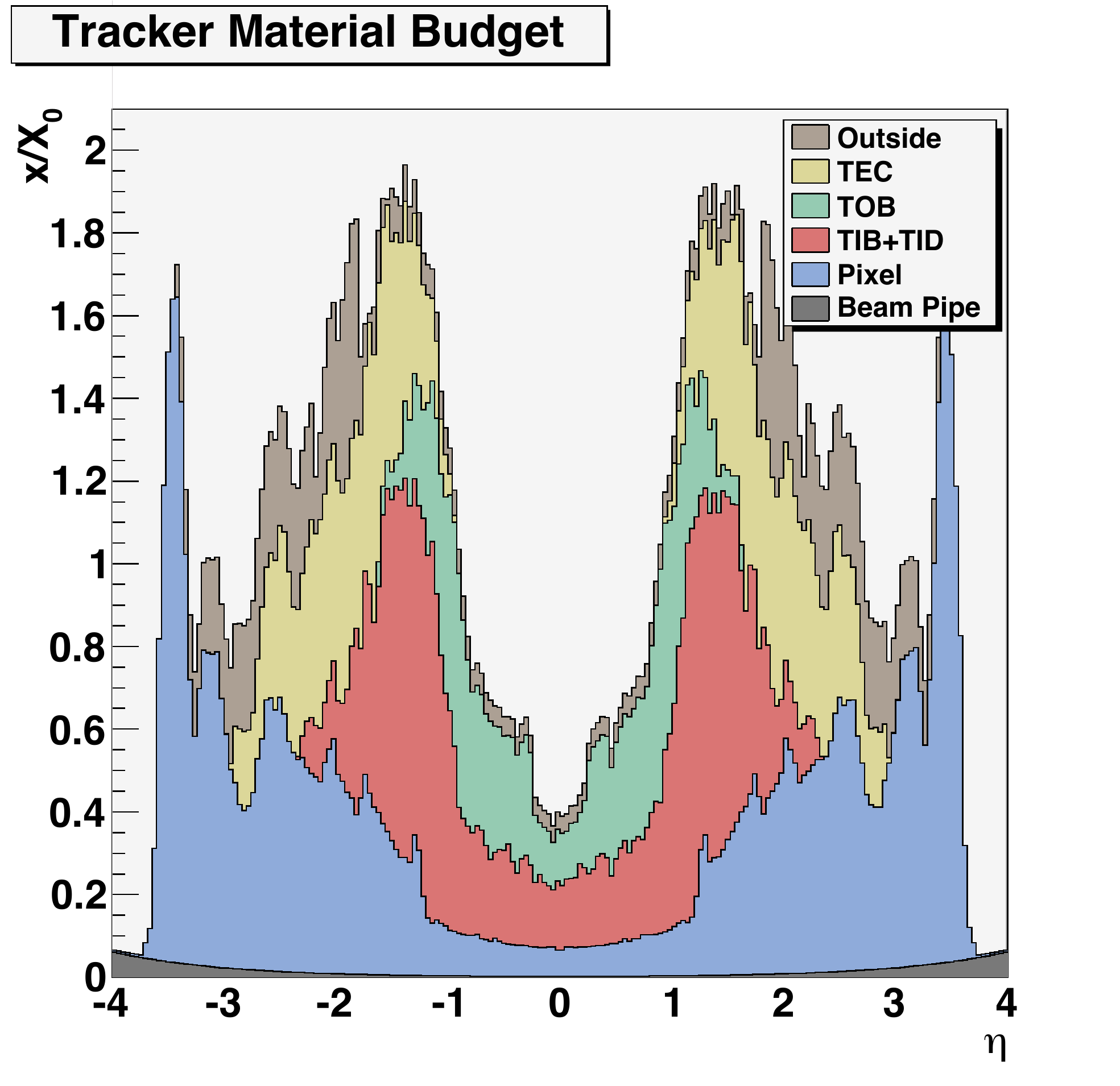}
\includegraphics[width=0.48\textwidth]{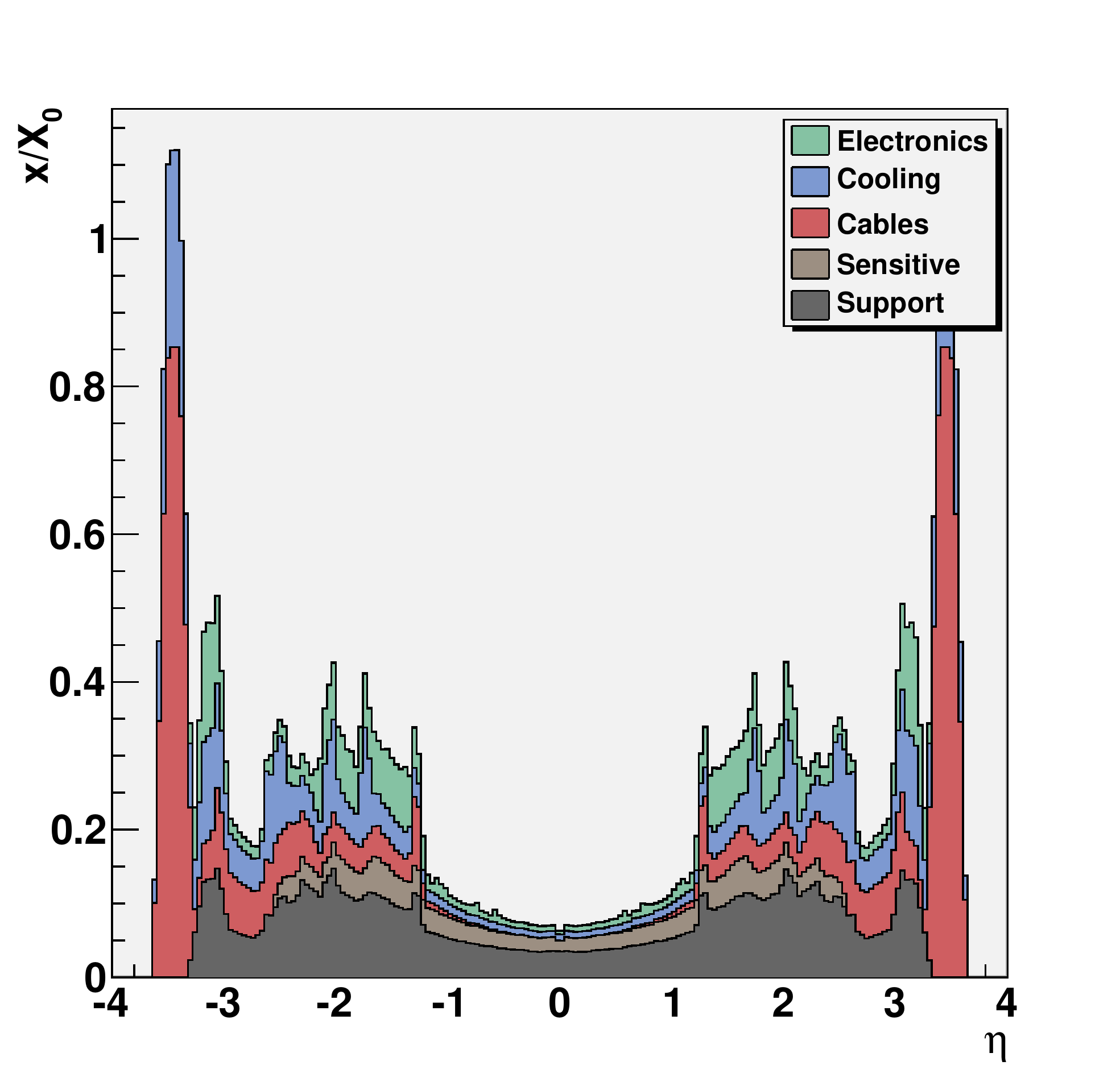}
\caption{Left: total integrated material budget of the CMS tracker in terms of radiation lengths $x/X_0$ as a function of pseudo-rapidity, including the forward pixel detector. The contributions of the various subdetectors are stacked.  Right: material budget for the pixel barrel detector only, showing the various categories of material. \label{fig:RadLength}}
\end{center}
\end{figure*}

Figure~\ref{fig:RadLength} shows the integrated material budget in terms of radiation lengths  for all components of the CMS tracker, including the pixel detector. 
The main pixel contribution to the material budget is in the region at $|\eta| > 1.2$  and 
is due to the material in the endflange and the inner part of the  supply
tubes.
Therefore, future detector upgrades will concentrate on the 
reduction of the material in these components.

%%%%%\\\\\\\\\\\\\\\\\\\\\\\\\\\\\\\\\\\\\\\\\\\\
%%%%%%%\\\\\\\\\\\\\\\\\\\\\\\\\\\\\\\\\\\\\\\\\\\\

%%%%%\\\\\\\\\\\\\\\\\\\\\\\\\\\\\\\\\\\\\\\\\\\\
%%%%%%%\\\\\\\\\\\\\\\\\\\\\\\\\\\\\\\\\\\\\\\\\\\\

%%%%%\\\\\\\\\\\\\\\\\\\\\\\\\\\\\\\\\\\\\\\\\\\\
%%%%%%%\\\\\\\\\\\\\\\\\\\\\\\\\\\\\\\\\\\\\\\\\\\\

%%%%%\\\\\\\\\\\\\\\\\\\\\\\\\\\\\\\\\\\\\\\\\\\\
%%%%%%%\\\\\\\\\\\\\\\\\\\\\\\\\\\\\\\\\\\\\\\\\\\\


\begin{thebibliography}{99}
\bibitem{bib:detectorPaper} The CMS Collaboration, ``The CMS experiment at the CERN LHC'', J. Inst. {\bf 3} (2008) S08004.
\bibitem{bib:trackerTDR} The CMS Collaboration, ``Tracker Technical Design Report'', CERN/LHCC 98-6  (1998).
\bibitem{bib:trackerTDRAddendum} The CMS Collaboration, ``Addendum to the CMS Tracker TDR'', CERN/LHCC 200-016 (2000).
\bibitem{bib:NIMAoverview} H. C. K\"astli et al., ``CMS barrel pixel detector overview'', Nucl. Instrum. Meth. A {\bf 582} (2007) 724.
\bibitem{bib:NIMA582_776} S. K\"onig et al., ``Building CMS pixel barrel detector modules'', Nucl. Instrum. Meth. A {\bf 582} (2007) 776.
\bibitem{bib:SensorDesign} Y. Allkofer et al., ``Design and performance of the silicon sensors for the CMS barrel pixel detector'', Nucl. Instrum. Meth. A {\bf 584} (2008) 25. 
\bibitem{bib:ROCdesign} H. C. K\"astli et al., ``Design and performance of the CMS pixel detector readout chip'', Nucl. Instrum. Meth. A {\bf 565} (2006) 188.
\bibitem{bib:NIMA565_73} D. Kotlinski et al., ``The control and readout systems of the CMS pixel barrel detector'', Nucl. Instrum. Meth. A {\bf 565} (2006) 73. 
\bibitem{bib:NIMA565_62} S. K\"onig et al., ``Assembly of the CMS pixel barrel modules'', Nucl. Instrum. Meth. A {\bf 565} (2006) 62.
\bibitem{bib:laserWelding} laser welding: CREATECH AG, Gaswerkstrasse 67, CH-4901 Langenthal, Switzerland: {\verb|http://www.createch.ch|}
\bibitem{bib:Airex} ALCAN AIREX AG, CH-5643 Sins, Switzerland: {\verb|http://www.alcanairex.com|}
\bibitem{bib:GEANT4A} S. Agostinelli et al., ``GEANT4 - A Simulation Toolkit'', Nucl. Instrum. Meth. A {\bf 506} (2003) 250.
\bibitem{bib:GEANT4B} J. Allison et al., ``Geant4 developments and applications'', IEEE Transactions on Nuclear Science {\bf 53} (2006) 270.
\bibitem{bib:DDD} M. Case et al., ``Detector Description Domain Architecture and Data Model'', CMS Note {\bf 2001-057} (2001).
 \end{thebibliography}
\end{document}